\documentclass[12pt,aps,amsmath,latexsym,amsfonts,letterpaper]{JHEP3}
\usepackage[all]{xy}
\usepackage{epsfig}
\usepackage{subfigure}

\def\half{\textstyle{1\over2}}

\def\quarter{\textstyle{1\over4}}

\def\ie{{\it i.e.,}\ }

\newcommand{\be}{\begin{equation}}
\newcommand{\ee}{\end{equation}}
\newcommand{\bea}{\begin{eqnarray}}
\newcommand{\eea}{\end{eqnarray}}
\newcommand{\bml}{\begin{mathletters}}
\newcommand{\eml}{\end{mathletters}}
\newcommand{\al}{\ensuremath{\alpha}}

\newcommand{\ka}{\ensuremath{\kappa}}

\renewcommand{\tt}{\ensuremath{\textrm}}

\renewcommand{\d}{\ensuremath{{\rm d}}}
\newcommand{\del}{\ensuremath{\partial}}

\newcommand{\ba}{\begin{eqnarray}}
\newcommand{\ea}{\end{eqnarray}}

\newcommand{\lab}{\label}

\newcommand{\beq}{\begin{equation}}
\newcommand{\eeq}{\end{equation}}
\newcommand{\beqa}{\begin{eqnarray}}
\newcommand{\eeqa}{\end{eqnarray}}
\newcommand{\beqar}{\begin{eqnarray*}}
\newcommand{\eeqar}{\end{eqnarray*}}

\newcommand{\cit}{\cite}

\newcommand{\grad}{\ensuremath{\vec{\nabla}}}
\newcommand{\q}{\ensuremath{\vec{q}}}

\newcommand{\tony}{}
\title{No resonant tunneling in standard scalar quantum field theory}
\author{Edmund J. Copeland, Antonio Padilla and Paul M. Saffin\\
School of Physics and Astronomy, University Park, University of
Nottingham, Nottingham NG7 2RD, UK}

\date{\today}

\maketitle

\abstract{We investigate the nature of resonant tunneling in {\tony standard scalar} Quantum Field Theory. Following the pioneering work 
of Banks, Bender and Wu we describe the quantum field theory in terms of infinite dimensional 
quantum mechanics and utilize the ``Most probable escape path'' (MPEP) as the class of paths which dominate 
the path integral in the classically forbidden region. Considering a 1+1 dimensional field theory example we show 
that there are five conditions that any associated bound state in the classically allowed region must satisfy if 
resonant tunnelling is to occur, and we then proceed to show that it is impossible to satisfy all 
five conditions simultaneously. }

\preprint{arXiv:0709.0261 [hep-th]}

\begin{document}

\section{Introduction} 
\lab{intro}
One of the remarkable features of quantum mechanics is that it gives particles the ability to tunnel
to places that would not be allowed classically. By now the theory of tunneling 
is well developed and has been used in
many contexts, from the first applications in nuclear decay \cite{gamow:1928, condon:1928}
using quantum mechanics, to creating a Universe from nothing \cite{Coleman:1980aw} with quantum field
theory. Typically one finds that although a quantum particle can tunnel through a barrier the transmission
is exponentially damped, there are however circumstances where the barrier is effectively transparent
to particles of a particular energy. This is known as resonant tunneling.

One way to understand resonant tunneling is to use an analogue from classical electromagnetism 
(see for example \cite{eisberg-resnick}). Just as
quantum tunneling finds an electromagnetic counterpart in {\it frustrated total internal reflection},
so resonant tunneling has a counterpart in the Fabry-P\'{e}rot interferometer.
The Fabry-P\'{e}rot device consists of two parallel, partially silvered mirrors, such that incident
light may either pass directly through the device or suffer a number of reflections inside the
cavity. If the width of the cavity is given by a half-integer number of wavelengths, then the phase of
a ray leaving the device does not depend on the number of reflections inside the cavity and so will
constructively interfere.
A rather direct comparison with quantum mechanics
may be seen by studying the phenomenon using path integrals \cite{Zhota:1990}.
From this perspective we consider the path integral of the classical paths through a double-barrier
system (e.g. Fig. \ref{fig:pot2}), where the path in the classically forbidden region is replaced by classical paths using
imaginary time. The quantum phase of the particle is given by the action angle variable,
so that if the central region between the barriers is half a de Broglie wavelength then the multiple
classical paths in the path-integral will constructively interfere giving resonant tunneling. The first
observation of resonant tunneling was that of Chang {\it et al} using semiconductors \cite{chang:1974,mizuta}.

It is worth stressing that the condition on the action angle variable for resonant tunneling (\ref{eq:BohrSommerfeld})
is precisely that required for a bound state to exist in the central region of Fig. \ref{fig:pot2}.
Put another way, resonant tunneling {\em relies} on there being a
bound state. One then finds that particles whose energy matches that of the bound state may 
tunnel through the barriers with probability close to unity. So, if we are to find resonant quantum tunneling
in field theory, a minimum requirement is that there is some bound state which the system can use as a
springboard to reach the true vacuum. The question then is what should such a bound state look like? We shall argue that
the natural choice for such a bound state, at least in standard  scalar quantum field theory\footnote{By a ``standard'' scalar field, we mean one with a canonical kinetic term.}, is the oscillon
\cite{bbm,Bogolyubsky:1976yu,Bogolyubsky:1976nx,Ventura:1976va,Gleiser:1993pt,Copeland:1995fq,Saffin:2006yk}
a lump-like configuration
of the scalar field whose amplitude varies in time. We shall also argue that the requisite oscillon 
that would facilitate tunnelling from the false vacuum does not exist.

The motivation for the present study comes from an observation of Tye \cite{Tye:2006tg} concerning the myriad
vacua in the string landscape. Tye argues that, owing to the huge number of vacua, there is likely to be one direction
which satisfies some criterion for resonant tunneling, in which case we would find ourselves preferentially in
a Universe at the end of a chain of resonant tunneling events. {\tony Given that quantum mechanics can be recovered from
quantum field theory in the homogeneous limit, perhaps it is natural to expect resonant tunneling to occur in field theory.  However, one can immediately see the limitations of imposing  spatial homogeneity since this would not allow any kind of  bubble  nucleation! The problem lies in the fact that the spatially homogeneous configurations represent a measure zero subset of the full configuration space, even for finite spatial volume.} It is more
appropriate to picture quantum field theory as infinite-dimensional quantum mechanics, using the formalism of \cite{Banks:1973ps},
but then we must take care when applying our intuition from quantum mechanics.

We shall start in section \ref{tunQM} with a reminder of tunneling in quantum mechanics, using the WKB aproximation.
We extend this discussion in section \ref{resQM} to cover resonant tunneling in one dimension, and then $N$-dimensions
in section \ref{WKNQM}. The formalism for $N$-dimensional quantum mechanics is then generalized to field theory
in section \ref{WKBQFT}, and then we go on to discuss resonant tunneling in quantum field theory and prove a no-go
theorem. {\tony For illustrative purposes, we present a thin wall analysis in section \ref{sec:twl} before finishing with our conclusions}.

\section{Tunneling in quantum mechanics} 
\lab{tunQM}
Let us begin with a review of tunneling in quantum mechanics (closely following ~\cit{Tye:2006tg, Merz,froman}). 
Consider a particle of mass $m$ moving in a one dimensional potential, $V(q)$. Quantum mechanically, the 
particle is described by its wavefunction, $\psi(q)$, satisfying the time independent Schrodinger equation,
\be
-\frac{\hbar^2}{2m} \frac {d^2 \psi}{dq^2}+V(q) \psi=E \psi.
\ee
Provided that the WKB approximation is valid, the wavefunction (in the semi classical limit) is given by
\be
\psi(q) \cong \frac{\al_+}{\sqrt{k(q)}} \exp\left[ \frac{i}{\hbar} \int^q dq' k(q')\right]
             +\frac{\al_-}{\sqrt{k(q)}} \exp \left[ -\frac{i}{\hbar} \int^q dq' k(q')\right], 
\ee\be
k(q)=\sqrt{2m(E-V(q))},
\ee
in the classically allowed region, $E>V(q)$. We see that it is composed of a positive momentum 
piece ($\alpha_+$) and a negative momentum piece ($\alpha_-$).
In the classically forbidden region, $E<V(q)$, we have 
\be
\psi(q) \cong \frac{\beta_+}{\sqrt{\ka(q)}} \exp \left[ \frac{1}{\hbar} \int^q dq' \ka(q')\right]
             +\frac{\beta_-}{\sqrt{\ka(q)}} \exp \left[ -\frac{1}{\hbar} \int^q dq' \ka(q')\right], 
\ee\be
\ka(q)=\sqrt{2m(V(q)-E)},
\ee
Now suppose we wish to tunnel between two classically 
allowed regions, I and III, separated by a classically forbidden region, II,  as shown in Fig. \ref{fig:pot1} 

\FIGURE{\centerline{\includegraphics[width=8cm, height=8cm]{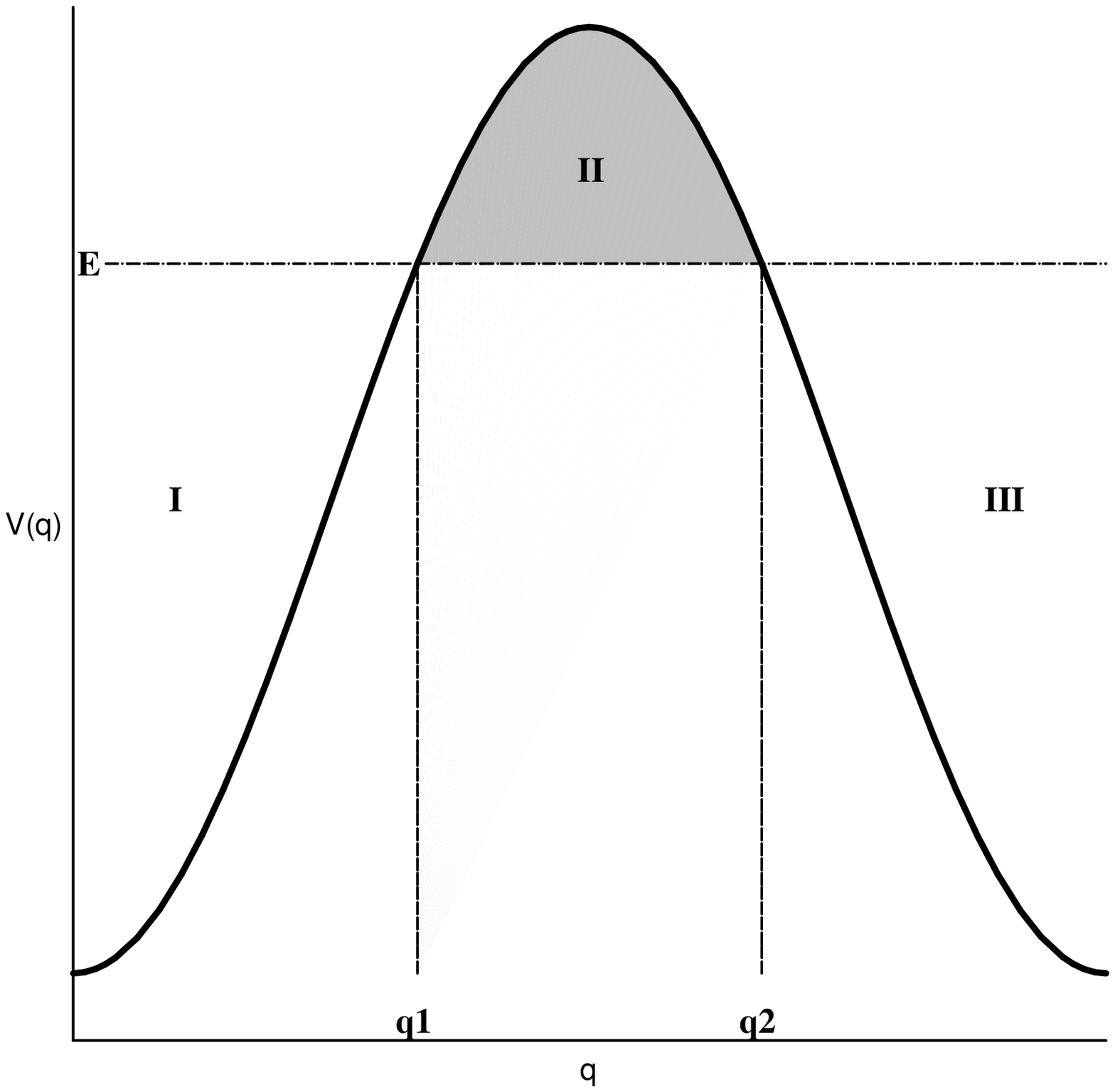}}
\caption{A single potential barrier, II, separating two classically allowed regions, I and III,
         for a particle of energy $E$.} \label{fig:pot1}}

By matching the general  solution in the forbidden region, II, onto the adjacent regions, I and III, 
it can be shown using the WKB connection formulae (appendix \ref{appendixA}) 
that the transmission coefficient 
for a particle incident on the left of the barrier is given by
\be
T_{ I \to III}=\left\vert \frac{\al_+^{III}}{\al_+^I}\right\vert^2=4/\Theta^2,
\ee
where
\be
 \Theta=2 \exp\left[\frac{1}{\hbar}\int_{q_1}^{q_2} dq' \ka(q')\right].
\ee
Note that  $q_1$ and $q_2$ are the classical turning points. 
Typically, $\Theta \gg 1$, so $T_{I \to III} \ll 1$. The probability of tunneling through 
the barrier, II,  is therefore  exponentially suppressed\footnote{This is sometimes quoted as 
$T_{ I \to III}=\left\vert \frac{\al_+^{III}}{\al_+^I}\right\vert^2=4\left( \Theta+\frac{1}{\Theta}\right)^{-2}$
~\cite{Merz}\cite{xu:1993}, however, as noted in appendix \ref{appendixA}, the WKB approximation
cannot consistently predict the $1/\Theta$ term \cite{froman,llqm}.}.

\section{Resonant tunneling in quantum mechanics} \lab{resQM}
We will now demonstrate how resonant tunneling can occur in quantum mechanics. For this we need three distinct 
classically allowed regions, separated by classically forbidden regions as shown in Fig. (\ref{fig:pot2}). 
There are now four turning points given by $q_1$, $q_2$, $q_3$ and $q_4$.

\FIGURE{\centerline{\includegraphics[width=8cm, height=8cm]{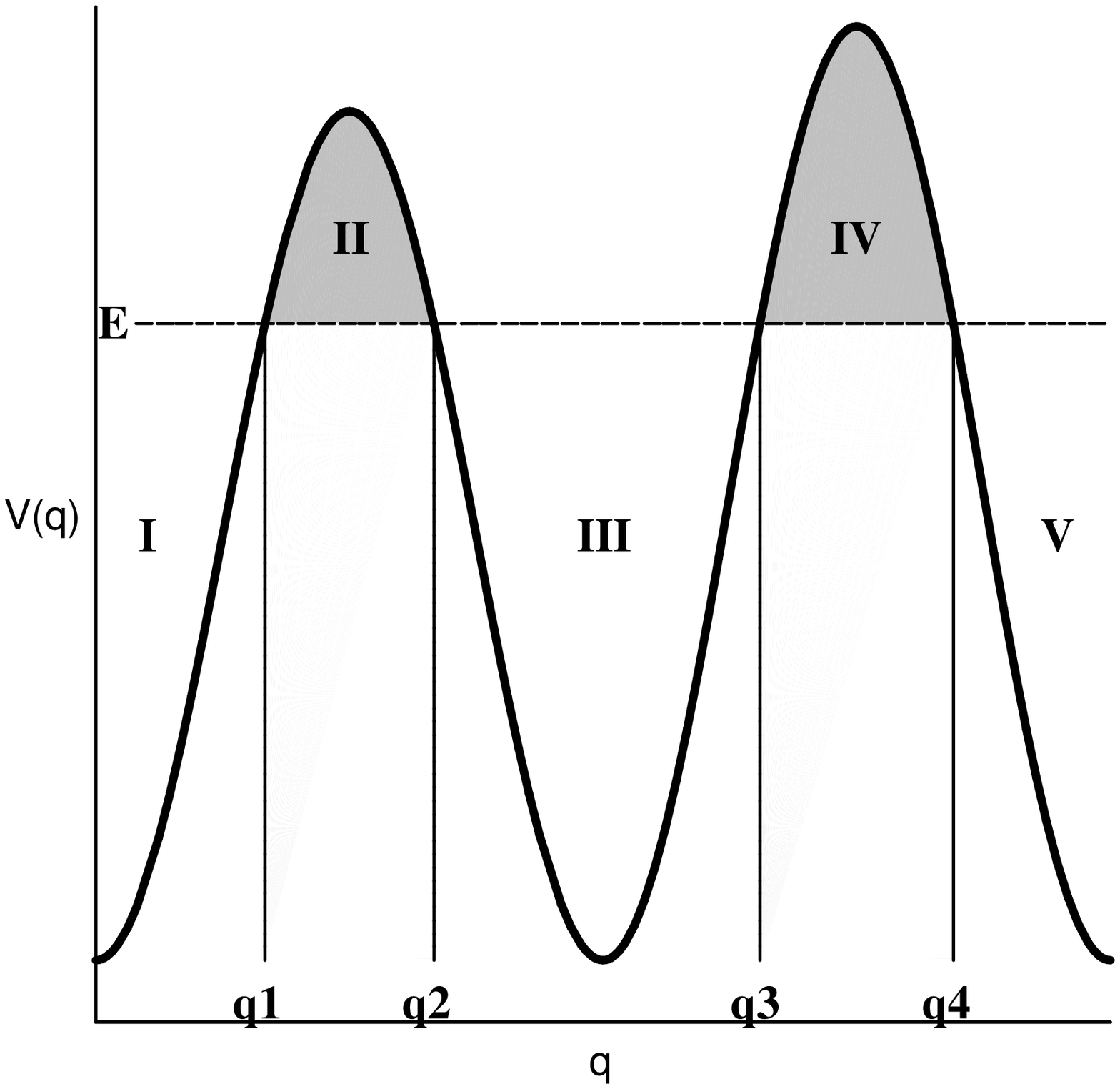}}
\caption{Two potential barriers, II and IV, separating three classically allowed regions, I, III and V,
         for a particle of energy $E$.} \label{fig:pot2}}

For $q \gg q_4$, the wavefunction is given by the following
\be
\psi(q) \cong \frac{\al_+^{V}}{\sqrt{k(q)}} \exp\left[\frac{i}{\hbar}\int^q_{q_4}dq'~k(q')\right] 
+ \frac{\al_-^{V}}{\sqrt{k(q)}} \exp\left[-\frac{i}{\hbar}\int^q_{q_4}dq'~k(q')\right].
\ee
Again, using the WKB connection formulas, it can be shown that
the tunneling amplitude is given by
\ba
T_{ I \to V}=\left\vert \frac{\al_+^{V}}{\al_+^I}\right\vert^2
=4/(\Theta\Phi\cos W)^2,
\ea
where 
\be
W=\frac{1}{\hbar}\int_{q_2}^{q_3} dq' ~k(q'), \qquad \Phi=2 \exp\left[\frac{1}{\hbar}\int_{q_3}^{q_4} dq' \ka(q')\right].
\ee

The condition we require for resonant tunneling to occur is that
\be
\label{eq:BohrSommerfeld}
W=(n+1/2)\pi, \qquad n \in \mathbb{Z}, \lab{bs}
\ee
in which case we see that the transmission rate diverges. Physically, of course, $T_{I\rightarrow V}$ is bounded
to be less than unity and there is another term which becomes important that the WKB analysis strictly cannot give.
If we were to, illegally, use the connection formulae without taking account of their directional nature one would
find the oft-quoted result 
\ba
T_{ I \to V}=\left\vert \frac{\al_+^{V}}{\al_+^I}\right\vert^2
=4\left[ \left(\Theta\Phi+\frac{1}{\Theta\Phi}\right)^{2}\cos^2 W+\left(\Theta/\Phi+\Phi/\Theta\right)^2\sin^2 W \right]^{-1},
\ea
showing how an extra term comes in as $\cos(W)\rightarrow 0$.

To interpret (\ref{eq:BohrSommerfeld}) we note that this is precisely the Bohr-Sommerfeld
quantization condition, a requirement for the existence of a bound state in the central, classically
allowed region; it is a statement that the allowed region is equal in width to a half-integer number of de Broglie wavelengths.
The picture we then have is that this bound state corresponds to a particle which oscillates
in the classically allowed region between turning points $q_2$ and $q_3$. As it oscillates it picks up a quantum phase,
and if that phase satisfies (\ref{eq:BohrSommerfeld}) then all such paths in the path-integral will constructively interfere,
leading to resonant tunneling.
%

\FIGURE{\centerline{\includegraphics[width=8cm, height=8cm]{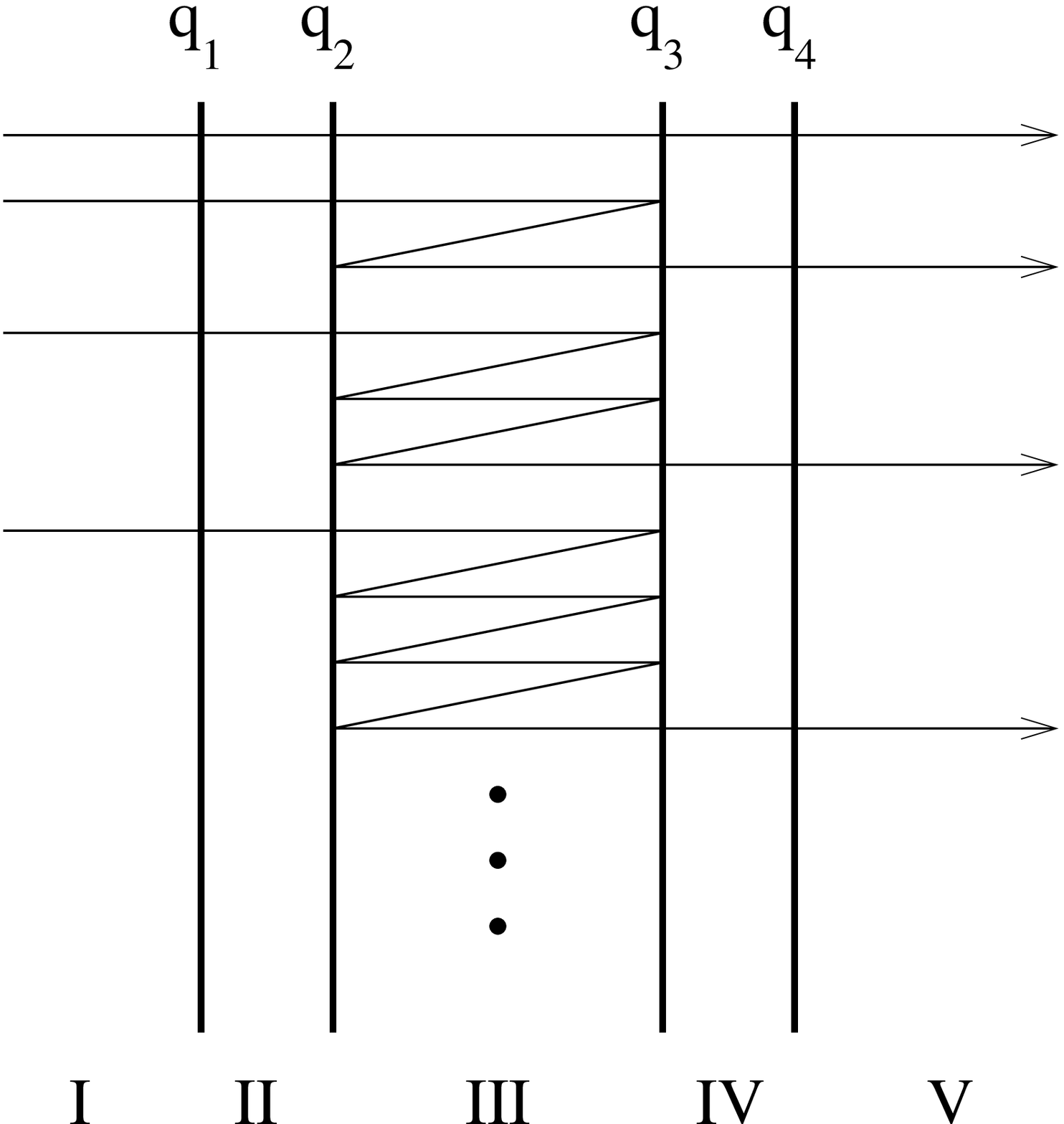}}
\caption{A plot showing different classical paths for the path-integral, these interfere constructively
        in resonant tunneling.} \label{fig:resPhase}}

There are two key points to take away from this semi-classical analysis:
\begin{itemize}
\item
the existence of a classical solution which oscillates between the stationary points.
\item
the quantum phase which such an oscillating solution acquires is $(n+1/2)\pi$.
\end{itemize}

\section{The WKB approximation in $N$-dimensional quantum mechanics} \lab{WKNQM}
In order to investigate the possibility of resonant tunneling in quantum field theory, we will need to develop the analogue 
of the WKB approximation. To do this we begin with a study of quantum mechanics in $N$-dimensions~\cite{Banks:1973ps} before extending 
our results to field theory~\cite{BC}. The leap from quantum mechanics in $N$ dimensions to field theory is achieved by 
carefully taking the limit as $N \to \infty$, as we will discuss later.

Consider  the mechanics of a particle of unit mass in $N$-dimensions. The classical path of the particle  is  given by 
$\vec{q}(t)=(q_1(t), \ldots, q_N(t))$, and is found by extremizing the action
\be
S=\int dt~ \left[\half \dot{\vec{q}} \cdot \dot{\vec{q}} -V(\vec{q})\right], \lab{actionN}
\ee
Quantum mechanically, a particle, of energy $E$ is described by the wavefunction $\psi(\vec{q})$ satisfying the  time 
independent Schrodinger equation, $\hat H \psi=E\psi$. To derive the precise form of the Hamiltonian operator, $\hat H$, 
we first need to derive the Hamiltonian for the action (\ref{actionN}). This is given by
\be
H=\frac{1}{2} \vec{p} \cdot \vec{p}+V(\vec{q}), \lab{hamN}
\ee
where $\vec{p}=\dot{\vec{q}}$ is the conjugate momentum. By letting $\vec{p} \to -i\hbar \vec{\nabla}$,  we promote 
the Hamiltonian (\ref{hamN}) to an operator, $H \to \hat H$, and Schrodinger's equation (in $N$ dimensions) becomes
\be
\left[-\frac{\hbar^2}{2} \vec{\nabla}^2+V(\vec{q})\right] \psi=E \psi. \lab{schroN}
\ee
As in 1-dimensional quantum mechanics, let us recast the wavefunction as the exponential of another  function 
$\sigma(\vec{q})$, as follows
\be
\psi(\vec{q})=e^{-\sigma(\vec{q})/\hbar}, \qquad \sigma(\vec{q})=\sigma_0(\vec{q})+\hbar \sigma_1(\vec{q})+\mathcal{O}(\hbar^2)
\ee
Using this semi-classical approximation, Schrodinger's equation gives
\ba
\vec{\nabla} \sigma_0 \cdot \vec{\nabla} \sigma_0 &=& 2(V-E) \lab{wkb1},\\
\vec{\nabla} \sigma_0 \cdot \vec{\nabla} \sigma_1 &=& \vec{\nabla}^2 \sigma_0/2. \lab{wkb2}
\ea
Clearly if $\sigma(\q)$ is real, we must be in the classically {\it forbidden} region ($E<V$). Let us assume for the moment 
that this is indeed the case. In one dimension, equations (\ref{wkb1}) and (\ref{wkb2}) have a straightforward solution. 
In more than one dimension this is no longer the case owing to the ambiguity in the direction of the gradient $\vec{\nabla}$. 
This problem was resolved by Banks, Bender and Wu by introducing the notion of the ``most probable escape path'' 
(MPEP)~\cite{Banks:1973ps}. The basic idea is 
that the path integral is dominated by the contribution from a discrete set of paths. As is well known, in the classically 
allowed region, these paths correspond to the classical solutions. However, in the classically forbidden region, they  
correspond to the MPEPs. By projecting the WKB equations (\ref{wkb1}) and (\ref{wkb2}) onto a MPEP, we are able to reduce the 
dimensionality of the problem. This enables us to (approximately) solve for the wavefunction close to a MPEP. Contributions to 
the wavefunction from fluctuations away from the MPEPs are subdominant.

The formal definition of a MPEP is as follows. Consider  a curve,  $\vec{Q}(\lambda) \in \mathbb{R}^N$, parametrized by 
$\lambda$. The curve  has tangent vector $\vec v_\parallel(\lambda)=\del \vec{Q} /\del \lambda$, and $N-1$ orthogonal 
normal vectors $\vec v_\perp^{~i}(\lambda),~i=1, \ldots, N-1$, satisfying
\be
\vec v_\parallel \cdot \vec v_\perp^{~i}=0, \qquad \vec v_\perp^{~i} \cdot \vec v_\perp^{~j} \propto \delta^{ij}.
\ee
The gradient operator on the curve, $\grad \vert_{\q=\vec{Q}}$, can also be expanded as follows
\be
\grad \vert_{\q=\vec{Q}}=\frac{\vec v_\parallel}{|\vec v_\parallel|^2}\left(\vec v_\parallel \cdot \grad \right)\vert_{\q=\vec{Q}}
+\sum_{i=1}^{N-1}\frac{\vec v_\perp^{~i}}{|\vec v_\perp^{~i}|^2}\left(\vec v_\perp^{~i} \cdot \grad \right)\vert_{\q=\vec{Q}}.
\ee
Of course, everything we have said thus far would be true for {\it any} curve in $\mathbb{R}^N$. What singles out the 
MPEP is the condition
\be
\vec v_\perp^{~i} \cdot \grad \sigma\vert_{\q=\vec{Q}}=0, \qquad i=1,\ldots, N-1.\lab{MPEPcondition}
\ee
We will see later that this path corresponds to a stationary solution to  the Euclidean action,  and as such gives the 
dominant contribution to the  path integral. 
In practice, what this means is that the wave function is peaked along the MPEP.
In order to solve equations (\ref{wkb1}) and (\ref{wkb2}) along the MPEP, 
we reparametrize it as $\vec{Q}(\lambda(s))$, where we have introduced, $s$, the proper distance along the curve, satisfying
\be
\frac{d}{ds}=\frac{\vec v_\parallel}{|\vec v_\parallel|} \cdot \grad\vert_{\q=\vec{Q}}  
\qquad \Longrightarrow 
\qquad ds=|d\vec{Q}|=\sqrt{\frac{d\vec{Q}}{\d \lambda} \cdot \frac{d\vec{Q}}{\d \lambda}} d\lambda 
=|\vec v_\parallel|d\lambda.
\lab{ds}
\ee
Making use of the condition (\ref{MPEPcondition}),  we easily obtain the following solution to (\ref{wkb1}) and 
(\ref{wkb2}) on the MPEP:
\ba
\sigma_0(\vec{Q})&=&\pm\int^s ds~\sqrt{2\left(V(\vec{Q})-E\right)}, \lab{sig0}\\
\sigma_1 (\vec{Q}) &=& \quarter \ln\left[{2(V(\vec{Q})-E)}\right]. \lab{sig1}
\ea
Close to the MPEP, we can write
\be
\vec{q}\approx \vec{Q}(\lambda)+\sum_{i=1}^{N-1} a_i(\lambda) \vec{v}_\perp^i(\lambda) 
\ee
where we expect the coefficients $a_i(\lambda)$ to satisfy $|a|\lesssim{\cal O}(\hbar)$. 
Given the condition (\ref{MPEPcondition}), this enables us to extend the solutions (\ref{sig0}) and (\ref{sig1}) to a 
neighbourhood of the MPEP, by simply sending $\vec{Q} \to \vec{q}$.

In the semi-classical approximation, the wavefunction in the forbidden region ($E<V$) is dominated by its value close 
to the MPEP, where it is given by
\be
\psi(\vec{q}) \cong \frac{1}{\left[2(V(\vec{q})-E)\right]^{\quarter}} 
\left[ \beta_+e^{\frac{1}{\hbar} \int^s ds~\sqrt{2(V(\vec{q})-E)}}
      +\beta_-e^{ -\frac{1}{\hbar} \int^s ds~\sqrt{2(V(\vec{q})-E)}} \right], 
\ee
The problem has now been reduced to one of finding the MPEP itself. To do this we make use of its formal definition 
(\ref{MPEPcondition}), which in practice says that to leading order, $\sigma$ is stationary in the direction normal to 
the path. Let us consider an arbitrary fluctuation, $\delta \vec{Q}$ normal to the path  ($\delta \vec{Q} \cdot \vec{Q}=0$). 
After a brief amount of algebra, it can be shown by using (\ref{ds}) in (\ref{sig0}) and then varying, that
$\sigma_0(\vec{Q}+\delta \vec{Q})$ vanishes to first order in  $\delta \vec{Q}$, if and only if
\be
\frac{d^2 \vec{Q}}{d\lambda^2}-\grad V=0,  \lab{Xlamlam}
\ee
where we have chosen $\lambda$ such that 
\be
\left(\frac{ds}{d\lambda}\right)^2=2(V-E).
\ee
Note that (\ref{Xlamlam}) corresponds to the equation of motion derived from the Euclidean version of (\ref{actionN}), 
with $\lambda$ playing the role of {\it imaginary} time.

When the particle enters the classically allowed region ($E>V$), we repeat the above analysis with the MPEP replaced by the 
classical path, $\vec{Q}_{cl}(t)$, satisfying
\be
\frac{d^2 \vec{Q}_{cl}}{dt^2}+\grad V=0  \lab{Xtt}
\ee
Here $t$  plays the role of {\it real} time, and is related to the proper distance along  the curve by
\be
\left(\frac{ds}{dt}\right)^2=2(E-V)
\ee
In the semi-classical approximation,  the wavefunction close to  the classical path is given by
\be
\psi(\vec{q}) \cong \frac{1}{\left[2(E-V(\vec{q}))\right]^{\quarter}} 
                  \left[ \alpha_+e^{\frac{i}{\hbar} \int^s ds~\sqrt{2(E-V(\vec{q}))}}
                        +\alpha_-e^{ -\frac{i}{\hbar} \int^s ds~\sqrt{2(E-V(\vec{q}))}} \right], 
\ee

\section{The WKB approximation in quantum field theory} \lab{WKBQFT}
We will now generalise the results of the previous section to {\tony standard scalar} quantum field theory in $1+1$ dimensions, closely 
following~\cite{BC} {\tony (see also~\cite{devega})}. Such ideas are well established and have been used successfully to derive tunneling rates in
quantum field theory \cite{Coleman}.
One can easily generalize this to any number of dimensions.
Consider the {\tony standard} theory of a  scalar field, $\phi(t, x)$, evolving in time through a spatial volume, 
$\mathcal{V}$, under the infuence of a potential, $V(\phi)$. This is described by the action
\be
S=\int dt \int_\mathcal{V} dx ~\left[\half\dot\phi^2-\half \phi'^2-V(\phi)\right], \lab{action}
\ee
where $~\dot{}=\del/\del t$, and $~'=\del/\del x$. We can think of the field $\phi(t, x)$ as describing a quantum  
mechanical system in infinite-dimensional space, like so
\be
\{\phi(t, x), x \in \mathcal{V}\}=\{\phi(t, x_1), \phi(t, x_2), \ldots\}.
\ee
Thinking of the field in this way makes it easy to extend the  results of the previous section. In particular,  
functions of the vector, $\vec{x}$, will be replaced by functionals of the scalar field, $\phi$, and the scalar 
product  will be replaced by an integral over space. To illustrate this explicitly, consider the Hamiltonian derived 
from the action (\ref{action}). This is given by
\be
H=\int_\mathcal{V} dx~ \left[\half \pi^2~+\half \phi'^2+V(\phi)\right]. \lab{ham}
\ee
where the conjugate momentum $\pi=\dot\phi$. Comparing the field theory Hamiltonian (\ref{ham}) with the $N$-dimensional 
Hamitonian (\ref{hamN}), we see that we should make the following generalisations:
\ba
 \half {\vec{p}} \cdot {\vec{p}}\qquad &\to& \qquad  \int_\mathcal{V} dx~ \half \pi^2,
\\
\label{eq:Uphi}
V(\vec{q}) \qquad &\to& \qquad U[\phi]=\int_\mathcal{V} dx~ \left[\half \phi'^2+V(\phi)\right],
\ea
where we have introduced the generalised potential, $U[\phi]$~\cite{BC}. 
It is the form of this potential, rather than $V(\phi)$, that determines the tunneling of the system. For
this reason it is difficult to apply one's intuition from the quantum mechanics potential $V(\vec{q})$ (which
is the naive analogue of $V(\phi)$) to the field theory situation. 
To see this explicitly, suppose we take the simple potential\footnote{This is not a tunneling potential, we just use it to make our point.}
\ba
\label{eq:qftpot}
V(\phi)&=&\lambda\phi^4.
\ea
and consider the case of a field theory in a finite box of size $L$, such that $\phi$ vanishes at the
boundary. By making the Fourier decomposition,
\ba
\phi(t,x)&=&\sum_{n=1}^\infty q_n(t)\sin(2\pi n x/L).
\ea
we are able to calculate the {\it generalised} potential (\ref{eq:Uphi}),
\ba
\label{eq:qmqftpot}
U[\phi]\equiv U(\{q_n\})=\frac{\lambda}{32}
\sum_{nmp}&&q_nq_mq_p[q_{n-m+p}+q_{-n+m+n}-q_{-n+m-p}-q_{n-m-p}-q_{n+m+p}\nonumber\\
&&-q_{-n-m+p}+q_{-n-m-p}+q_{n+m-p}]+\sum_n\frac{\pi^2}{L}n^2q_n^2.
\ea
The generalised potential in (\ref{eq:qmqftpot})  bears very little resemblance
to the field theory potential (\ref{eq:qftpot}). This clearly demonstrates why one should proceed with extreme care applying quantum mechanical intuition to a field theory setting. It is the generalised potential, $U[\phi]$, that is relevant in tunneling situations, and the one ``seen'' by the MPEP.  Even if
the field theory potential (\ref{eq:qftpot}) looks similar to Fig. \ref{fig:pot2}, we have no reason to expect that the system will
support resonant tunneling.

To derive the field theory analogue of the 
Hamiltonian operator we let $\pi \to -i\hbar \frac{\delta}{\delta \phi(x)}$, and obtain the generalised Schrodinger equation
\be
\left[-\frac{\hbar^2}{2}\int_\mathcal{V} dx~\frac{\delta^2}{\delta \phi(x)^2}+U[\phi]\right]\psi[\phi]=E\psi[\phi]
\ee
The wavefunction, $\psi$  is a functional acting on an appropriately chosen  ``configuration space''.  Whereas in 
$N$-dimensional quantum mechanics, the ``configuration space'' was simply $\mathbb{R}^N$,  here it is the space of 
real valued functions on $\mathcal{V}$, satisfying some boundary condition on $\partial \mathcal{V}$. The norm, 
$|\psi[\phi]|^2$,  therefore measures the probability density for a given configuration $\phi$.

As before, we make a semi-classical approximation
\be
\psi[\phi]=e^{-\sigma[\phi]/\hbar}, \qquad \sigma[\phi]=\sigma_0[\phi]+\hbar \sigma_1[\phi]+\mathcal{O}(\hbar^2),
\ee
and derive the field theory version of the WKB equations
\ba
\int_\mathcal{V} dx~ \left(\frac{\delta \sigma_0[\phi]}{\delta \phi(x)}\right)^2&=&2(U[\phi]-E), \lab{wkb1field}\\
\int_\mathcal{V} dx~ \frac{\delta \sigma_0[\phi]}{\delta \phi(x)}\frac{\delta \sigma_1[\phi]}{\delta \phi(x)} 
      &=&\frac{1}{2} \int_\mathcal{V} dx~\frac{\delta^2 \sigma_0[\phi]}{\delta \phi(x)^2} .\lab{wkb2field}
\ea
The forbidden region now corresponds to $E<U$. In this region we need the analogue of the MPEP. This should be the path 
in configuration space that dominates the path integral. We therefore take the MPEP to be a curve $\Phi(\lambda, x)$ in 
this space, parametrized by $\lambda$. At each point $x \in \mathcal{V}$, the curve has tangent vector, 
$v_\parallel(\lambda, x)=\del \Phi/ \del \lambda$, and a continuous set of orthogonal normal vectors, 
$v_\perp(\lambda, x; y)$, satisfying
\be
\int_\mathcal{V} dx~v_\parallel(\lambda, x) v_\perp(\lambda, x;y)=0, 
\qquad \int_\mathcal{V} dx~v_\perp(\lambda, x; y) v_\perp(\lambda, x;y')\propto \delta(y-y').
\ee 
As before, the gradient on this curve can be expanded as follows
\ba
\left.\frac{\delta}{\delta \phi(x)}\right\vert_{\phi=\Phi}&=&
      \frac{v_\parallel(\lambda, x)}{\int_\mathcal{V} dx' v_\parallel(\lambda, x')^2}
      \int_\mathcal{V} dx'~v_\parallel(\lambda, x')\left. \frac{\delta}{\delta \phi(x')}\right\vert_{\phi=\Phi}\nonumber\\
   &&+\int dy~\left[\frac{ v_\perp(\lambda, x; y)}{\int_\mathcal{V} dx' v_\perp(\lambda, x'; y)^2}
       \int_\mathcal{V} dx'~v_\perp(\lambda, x'; y)\left. \frac{\delta}{\delta \phi(x')}\right\vert_{\phi=\Phi} \right].~
\ea
The MPEP is chosen so that the wavefunction is stationary relative to fluctuations normal to the path, \ie
\be
\int_\mathcal{V} dx'~v_\perp(\lambda, x'; y)\left.\frac{\delta \sigma[\phi]}{\delta \phi(x')} \right\vert_{\phi=\Phi}=0, 
\qquad \forall y. \lab{FTMPEPcond}
\ee
In direct analogy with the previous section, this condition ensures that the path corresponds to the stationary solution 
to the Euclidean action, and as such dominates in the  path integral.  As before, the WKB equations (\ref{wkb1field}) 
and (\ref{wkb2field}) can be solved along the MPEP to give,
\ba
\sigma_0[\Phi]&=&\int^s ds~\sqrt{2\left(U[\Phi]-E\right)}, \lab{sig0ft}\\
\sigma_1 [\Phi] &=& \quarter \ln\left[{2(U[\Phi]-E)}\right], \lab{sig1ft}
\ea
where we have introduced, $s$, the proper distance along the path, defined as:
\be
\frac{d}{ds}=\frac{\int_\mathcal{V} dx~v_\parallel(\lambda, x)\left.\frac{\delta}{\delta \phi(x)}\right\vert_{\phi=\Phi}}
                  {\left[\int_\mathcal{V} dx~v_\parallel(\lambda, x)^2\right]^{\half}}  
           \qquad \Longrightarrow \qquad ds=\left[\int_\mathcal{V} dx~ \left(\frac{d\Phi}{d \lambda}\right)^2 \right]^{\half} 
                  ~d\lambda. \lab{dsft}
\ee
Close to the MPEP, we can write
\be
\phi(x)\approx \Phi(\lambda, x)+\int dy~a(\lambda; y) v_\perp(\lambda, x; y).
\ee
where $|a| \lesssim \mathcal{O}(\hbar)$. Given the condition (\ref{FTMPEPcond}), we can now extend the solutions (\ref{sig0ft}) and (\ref{sig1ft}) to a 
neighbourhood of the MPEP, by sending $\Phi \to \phi$. In the semi-classical approximation, the wavefunction in the 
forbidden region ($E<U$) is dominated by its value close to  the MPEP, where it is given by
\be
\psi[\phi] \cong \frac{1}{\left[2(U[\phi]-E)\right]^{\quarter}} 
           \left[ \beta_+e^{\frac{1}{\hbar} \int^s ds~\sqrt{2(U[\phi]-E)}}
                 +\beta_-e^{ -\frac{1}{\hbar} \int^s ds~\sqrt{2(U[\phi]-E)}} \right], \lab{psiMPEP}
\ee
Again, the problem has been reduced to finding the MPEP itself. In close analogy with the previous section, it can be shown 
that if $\lambda$ is chosen so that
\be
\left(\frac{ds}{d\lambda}\right)^2=2(U[\Phi]-E),
\ee
then the MPEP satisfies
\be
\frac{d^2 \Phi}{d \lambda^2}+\frac{d^2 \Phi}{dx^2}-V'(\Phi)=0. \lab{Philamlam}
\ee
Of course, equation (\ref{Philamlam}) corresponds to the equations of motion derived from the Euclidean version of 
(\ref{action}), with $\lambda$ playing the role of imaginary time.

In the classically allowed region ($E>U$), the MPEP is replaced by the classical path, $\Phi_{cl}(t, x)$, satisfying
\be
\frac{d^2 \Phi_{cl}}{d t^2}-\frac{d^2 \Phi_{cl}}{dx^2}+V'(\Phi_{cl})=0, \lab{Phitt} \lab{cleom}
\ee
where $t$ plays the role of real time, and is related to the proper distance along the curve by
\be
\left(\frac{ds}{dt}\right)^2=2(E-U[\Phi_{cl}]). \lab{ds/dt}
\ee
In the semi-classical approximation,  the wavefunction close to the classical path is given by
\be
\psi[\phi] \cong \frac{1}{\left[2(E-U[\phi])\right]^{\quarter}} \left[\alpha_+e^{\frac{i}{\hbar} \int^s ds~\sqrt{2(E-U[\phi])}}
    +\alpha_-e^{ -\frac{i}{\hbar} \int^s ds~\sqrt{2(E-U[\phi])}} \right]. \lab{psicl}
\ee
We have now shown that the WKB approximation, which is so successful in quantum mechanics, has a direct
analogue in quantum field theory. Given that, we are able to apply the formalism to investigate the possibility of resonant tunneling in field theory. Comparing the field theory wavefunctions (\ref{psiMPEP}) and (\ref{psicl}) with their quantum mechanical counterparts in section \ref{tunQM}, there seems little reason to expect that resonant tunneling cannot occur for a suitably chosen generalised potential, $U[\phi]$. However, in the next section we will show that another  crucial ingredient is always missing in field theory: the  existence of a suitable ``bound state''. 

\section{Resonant tunneling in field theory: a no go theorem} \lab{nogo}
In the previous section, we showed how quantum field theory could be reduced, in the semi classical limit,  to quantum mechanics 
along a preferred path in configuration space. The ``preferred paths'' correspond to most probable escape paths (MPEPs) in the 
classically forbidden regions, and classical paths in the classically allowed regions.   Along a MPEP, the wavefunction is 
given by equation (\ref{psiMPEP}), and along the classical path by equation (\ref{psicl}). In order to develop a field theory 
analogue of resonant tunneling as descibed in section \ref{tunQM}, we need to be able to match the two solutions onto one another. 
Whenever a MPEP can be smoothly joined onto a classical path (or vice-versa) at a classical turning point, then it is clear that 
the matching conditions will mirror those in quantum mechanics (see~\cite{Merz}). 

We are now in a position to ask the following: (i)  what is required for resonant tunneling to occur in field theory, and (ii) is 
this even possible? Let us address the first question. Consider the dynamics of a scalar field described by the action (\ref{action}). 
A generic potential, $V(\phi)$,  is shown in figure~\ref{fig:pot3} with a local minimum at $\phi=0$, corresponding to the false 
vacuum, and a global minimum at $\phi=\phi_1$, corresponding to the true vacuum\footnote{Without loss of generality, we have 
chosen the false vacuum to the lie at a point where the potential vanishes. In principle there may be any number of local minima 
between $\phi=0$, and $\phi=\phi_1$, although this is not shown in the figure.}.

\FIGURE{\centerline{\includegraphics[width=8cm, height=8cm]{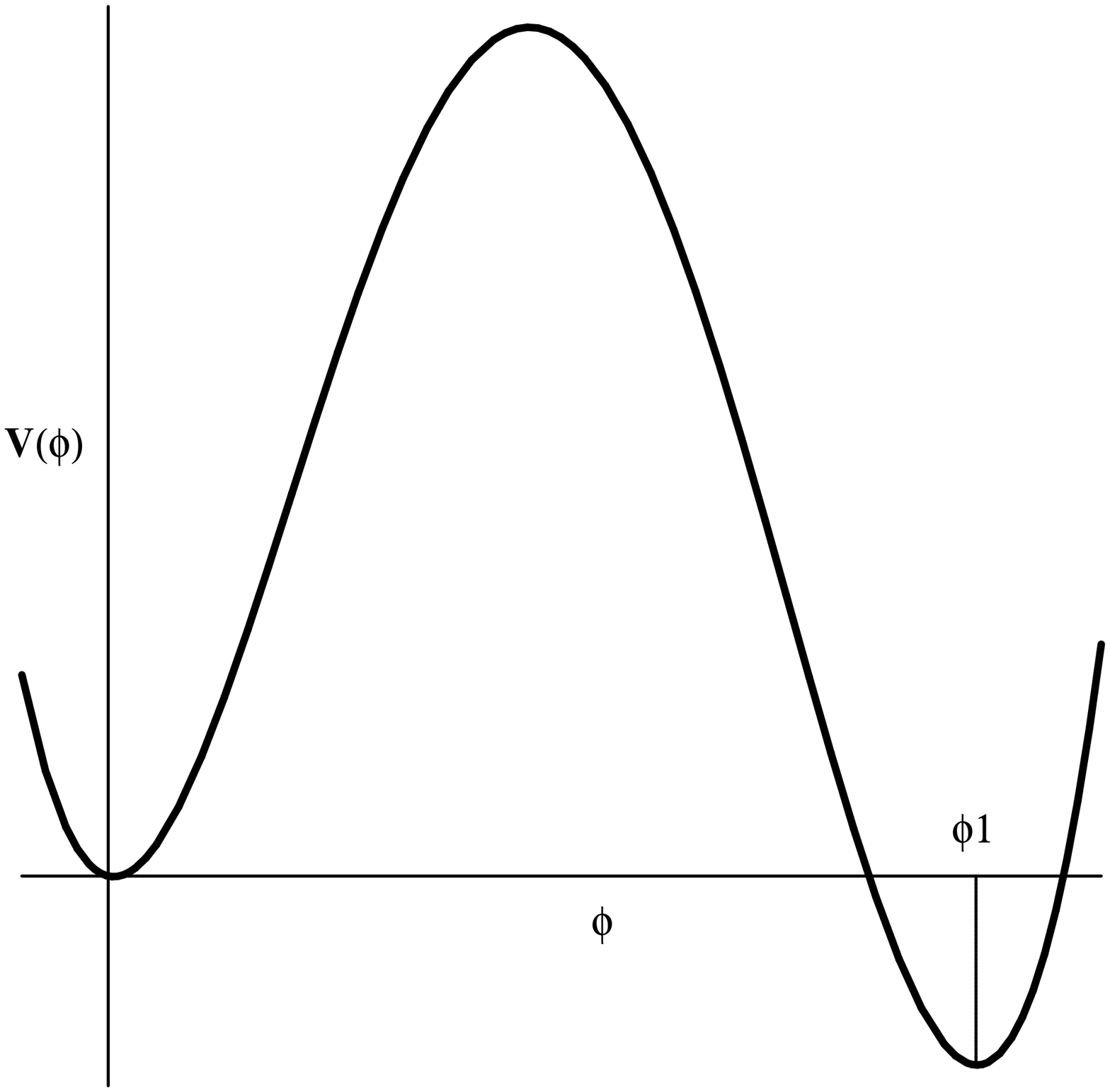}}
\caption{A generic potential, $V(\phi)$ with at least two minima,  at $\phi=0$ and $\phi=\phi_1$, 
         satisfying $V(\phi_1)<V(0)=0$.} \lab{fig:pot3}}

Now consider a tunneling process from false vacuum to true, through the nucleation of a bubble of true vacuum.  We can think 
of the false vacuum as a classical path $\phi=\Phi_{cl}^0(t, x)\equiv 0$. The expanding bubble is also a classical path 
$\phi=\Phi^1_{cl}(t, x)$, satisfying the same boundary conditions, \ie
\be
\left.\Phi\right\vert_{\partial \mathcal{V}}=\left.\frac{\del \Phi}{\del x}\right\vert_{\partial \mathcal{V}}=0, \lab{bc}
\ee
where $\partial \mathcal{V}$ refers to the spatial boundary which is usually taken to be at spatial infinity.  
To have any hope of tunneling between the two classical paths, they must have the same conserved energy.
\be
{E}=\int_{\mathcal{V}} dx~\left[
                \frac{1}{2} \left(  \frac{\del \Phi}{\del t}\right)^2+\frac{1}{2} \left(\frac{\del \Phi}{\del x}\right)^2
               +V(\Phi)            \right]
\ee
The energy of the false vacuum is, of course, zero. This is also true of the expanding bubble: the negative energy stored 
in the interior of the bubble being compensated for by the positive energy in the bubble wall~\cite{Coleman}. In a standard 
tunneling scenario, these classical paths are connected by a MPEP. The wavefunction $\psi[\phi]$ is damped exponentially 
as it follows the MPEP through the forbidden region, and so the probability of tunneling is suppressed. For {\it resonant} 
tunneling, analogous to that described in section \ref{tunQM}, we require the existence of another classical path,   
$ \Phi_{cl}^*(t, x)$. This must correspond to  some sort of ``bound state'', and be connected to the false vacuum by one MPEP, 
and to the expanding bubble by another.  {\tony We should emphasise that {\it resonant tunneling} from false vacuum to true is direct. It is {\it not} as if  we first tunnel to the bound state, stay there for a bit, and then tunnel through to the true vacuum. Of course, one could certainly imagine such a scenario, but this would simply correspond to two standard tunneling events, and would have nothing to do with resonant tunneling.} 

To understand what we need for a bound state we should reconsider what happens in quantum mechanics.  In the {\it classically allowed} region of Fig. \ref{fig:pot2}, there are classical paths {\it oscillating} between the turning points. These classical oscillating solutions qualify as bound states if they satisfy the Bohr-Sommerfeld quantization
condition (\ref{bs}). Provided such a bound state exists, resonant tunneling can occur if  the incoming particle has the same energy as the bound state.

In field theory we should therefore look for a classical, oscillatory solution which would play the role of the bound state
and act as the intermediary between the false and true vacuum. Moreover, this oscillatory solution must satisfy the field theory
analogue of (\ref{bs}), \ie
\be
\frac{1}{\hbar} \int_{s_1}^{s_2} ds~\sqrt{2(E-U[\Phi^*_{cl}])}=(n+1/2)\pi, \qquad n \in \mathbb{Z} \lab{bsft}
\ee
where $s$ is the proper distance along the curve, and $s_1$ and $s_2$ correspond to the classical turning points. Actually, 
it is convenient to replace the proper distance with real time, using equation (\ref{ds/dt}). Then (\ref{bsft}) is replaced by
\be
\frac{1}{\hbar} \int_{t_1}^{t_2} dt~2(E-U[\Phi^*_{cl}])=(n+1/2)\pi, \qquad n \in \mathbb{Z} \lab{bsft2}
\ee
where $t_1<t_2$ are the classical turning points in real time. Note that in addition, the momentum, $\del \Phi/\del t$, must 
vanish at these points. This enables us to connect the classical path  to a MPEP at one end ($t=t_1$), and to another MPEP, 
at the other ($t=t_2$). Crucially, this means that the classical solution, $ \Phi_{cl}^*(t, x)$ must be  stationary at {\it all} 
points in space,  on {\it at least two} separate occasions, $t_1$ and $t_2$. 

 All this considered, it is clear that for {\it resonant tunneling} to occur, the ``bound state''  $\Phi_{cl}^*(t, x)$ must satisfy 
each of the following conditions:
\begin{enumerate}
\item it is a solution to the classical field equations (\ref{cleom}),  other than the false vacuum.
\item it has zero energy 
\item it satisfies the boundary conditions (\ref{bc}) 
\item there exists $t_1, t_2 \in \mathbb{R}$, where $t_1<t_2$, such that 
      $\left.\frac{\del \Phi_{cl}^*}{\del t}\right\vert_{(t_1, x)}=\left.\frac{\del \Phi_{cl}^*}{\del t}\right\vert_{(t_2, x)}=0$, 
      $\forall x \in \mathcal{V}$.
\item it satisfies the ``bound state'' condition (\ref{bsft2})
\end{enumerate}
At first glance, it does not seem unreasonable that such a bound state could, in principle, exist. Oscillons, 
\cite{bbm,Bogolyubsky:1976yu,Bogolyubsky:1976nx,Ventura:1976va,Gleiser:1993pt,Copeland:1995fq,Saffin:2006yk}
for example, seem to satisfy most of the conditions: they are solutions to the field equations; they asymptote 
to the false vacuum; and they are everywhere stationary at certain times.
However, it is  not at all obvious that an oscillon will exist that has zero energy, or even that it satisfies 
the bound state condition.
There is a nice model with a logarithmic potential \cite{bbm,Ventura:1976va} for which the tunneling solution is
known analytically \cite{Ferraz de Camargo:1982sk}, moreover, the oscillon solutions are also 
simple to find (up to quadrature). In this case one finds that there are no zero energy oscillons.
Of course, this does not prove that such objects cannot exist, but it does help in our understanding of the problem.
In fact, we will now prove that there is no solution, oscillon or otherwise, that satisfies all five 
conditions.

Let us begin by assuming that a solution, satisfying all five conditions,  does in fact exist. 
Again, we work explicitly in 1+1 dimensions but the proof generalizes quite simply to higher dimensions.
One can easily check, that 
because of  first and fourth conditions, the integral, 
\be
I=\int_{t_1}^{t_2} dt~ \left[V(\Phi_{cl}^*)-\frac{1}{2} \left(  \frac{\del \Phi_{cl}^*}{\del t}\right)^2-\frac{1}{2} 
            \left(\frac{\del \Phi_{cl}^*}{\del x}\right)^2\right],
\ee
is constant everywhere in space.
We therefore evaluate it on the boundary, $\partial \mathcal{V}$, and use the third  condition to 
show that $I =0$, and therefore
\be
\int_{t_1}^{t_2} dt~ \left[\frac{1}{2} \left(  \frac{\del \Phi_{cl}^*}{\del t}\right)^2+\frac{1}{2} 
\left(\frac{\del \Phi_{cl}^*}{\del x}\right)^2\right]=\int_{t_1}^{t_2} dt~ V(\Phi_{cl}^*). \lab{Isol}
\ee
Now by the second condition, the solution has zero energy, from which it follows that
\be
\int_{t_1}^{t_2}dt\int_{\mathcal{V}} dx~\left[\frac{1}{2} \left(  \frac{\del \Phi_{cl}^*}{\del t}\right)^2+\frac{1}{2} 
     \left(\frac{\del \Phi_{cl}^*}{\del x}\right)^2
+V(\Phi_{cl}^*)\right]=0.
\ee
Reversing the order of integration, and making use of equation (\ref{Isol}), it is easy to show that
\be
\int_\mathcal{V} \int_{t_1}^{t_2} dt~ \left[ \left(  \frac{\del \Phi_{cl}^*}{\del t}\right)^2+ 
       \left(\frac{\del \Phi_{cl}^*}{\del x}\right)^2\right]=0.
\ee
It now follows that $\Phi_{cl}^* \equiv 0$, in other words, the required path has to be the false vacuum! 
But this cannot be, since it violates the first condition. Note that we have only used the first four 
conditions in arriving at this contradiction; the fifth, and final, condition was not required. 
This is simply an outcome of the fact that we have shown there is no classical path that 
has the required oscillatory property, so there is nothing for
the Bohr-Sommerfeld condition to pick from.
We conclude, therefore that there is no solution that can satisfy all five conditions. 

What does this mean practically? 
It means that, using the standard techniques introduced many years ago by Banks, Bender and 
Wu~\cite{Banks:1973ps}, there is no direct analogue in standard scalar quantum field theory, of  resonant tunneling in 
quantum mechanics. This may come as a surprise given that we were able to reduce the field 
theory problem to quantum mechanics using MPEPs and classical paths. However it was precisely 
the absence of a suitably oscillating classical path, or ``bound state'', that meant there 
simply were not enough ingredients available in field theory for an analogous resonance 
to occur. Tye has recently argued that a direct analogue of resonant tunneling in quantum 
mechanics may be relevant in the string landscape~\cite{Tye:2006tg}. Although we did not 
include gravity in our analysis, our results clearly {\tony cast doubts on the validity of this claim}. 

\section{Example: bubble nucleation in the thin wall limit} \lab{sec:twl}
In order to illustrate our result explicitly, let us consider the nucleation of a bubble of true vacuum, in the thin wall limit. As is well known~\cite{Coleman}, we can treat the bubble walls as membranes with an action given by
\ba
S_m&=&-\sigma\int{\rm d}^3\xi\sqrt{{\rm det}G}+\epsilon\int{\rm d}V{\rm d}t.
\ea
where $G$ is the induced metric on the world volume of the membrane, $\sigma$ the tension of the bubble wall,
$V$ the volume of the bubble and $\epsilon$ the difference in potential energy density between the inside and outside of the bubble.

If we impose spherical symmetry then the action of a single bubble wall leads to a Lagrangian for the
radius of the bubble given by
\ba
L&=&-4\pi\sigma R^2\sqrt{1-\dot{R}^2}+\frac{4\pi\epsilon}{3}R^3.
\ea
We can convert this to a Hamiltonian,
\ba
\lab{eq:ham}
H&=&\sqrt{P^2_R+(4\pi\sigma)^2R^4}-\frac{4\pi\epsilon}{3}R^3,\\\nonumber
 &=&P^2_R/(8\pi\sigma R^2)+...+4\pi\sigma R^2-\frac{4\pi\epsilon}{3}R^3,
\ea
where the $...$ signifies higher powers of the conjugate momentum, $P_R$. The important point to note is that the
potential term of (\ref{eq:ham}) takes the form indicated in Fig.~\ref{fig:pot4}. 
\FIGURE{\centerline{\includegraphics[width=8cm, height=8cm]{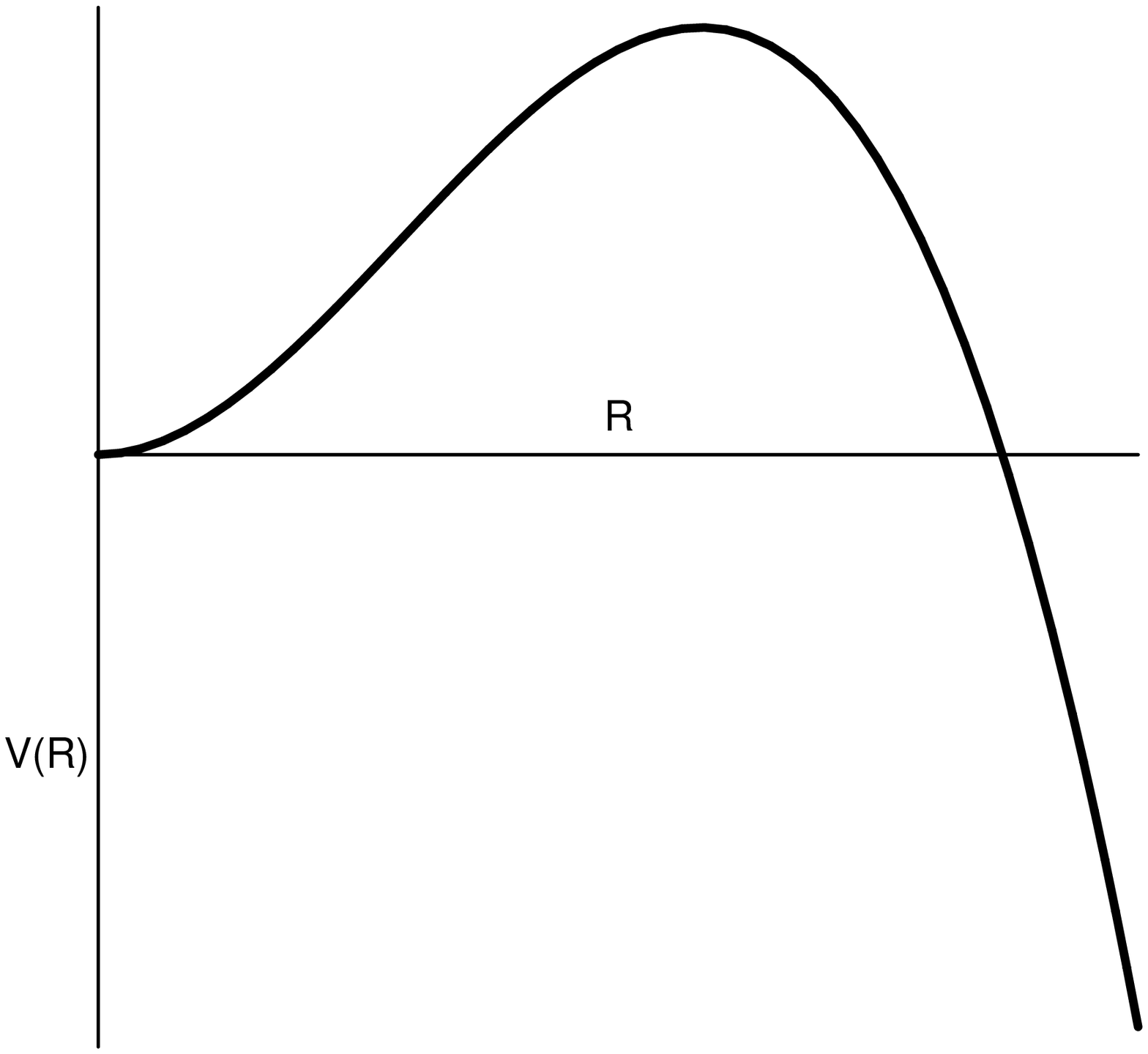}}
\caption{The potential for the position of the bubble wall, in the thin wall limit.} \label{fig:pot4}}
Using this figure we can understand
the nucleation of a single bubble as a quantum particle being described by (\ref{eq:ham}) which tunnels
from $R=0$ through the barrier, and then continues to roll down the potential, corresponding to the
bubble expanding.  To get {\it resonant tunneling} we need  a ``bound state'': a classical solution which is stationary on two separate occasions. In this thin wall set up, this would correspond to the bubble wall oscillating back and forth between two radii. It is easy to see that the potential does not have any region which 
could lead to such a bound state. Adding another bubble does not change this since in the thin wall limit the walls move independently of one another.

\section{Discussion} \lab{discuss}
Resonant tunneling through a double potential barrier is a well understood 
phenomenon in Quantum Mechanics, and can be described through the 
careful use of WKB techniques. For this resonance to occur, we require a number of key ingredients. The first is the existence of 
a classical bound state in the central classically allowed region. This is a 
classical solution that oscillates between the classical turning points, with 
a quantum phase of (n+1/2)$\pi$. If the energy of a wave incident on the barrier 
exactly matches the bound state energy, the wave will tunnel through the  double 
barrier with almost unit probability. This effect has been seen to occur in experiments 
involving semiconductors~\cite{chang:1974,mizuta}.

Given that a limit of quantum field theory is quantum mechanics, it is natural 
to expect that we should also be able to recover the resonance phenomena in the context of 
field theory. This would have a number of important consequences for cosmology, one of which was discussed in 
the recent work of Tye  \cite{Tye:2006tg}.  The cosmic landscape is made up of a huge number of vacua, which 
would mean there could be a direction which satisfies the criteria of resonant tunneling. If this were the 
case it would be possible to argue that we  have found ourselves in a Universe with such a small vacuum energy 
today because of resonant tunneling. 
However, in this paper, we have argued that there is no direct analogue of 
quantum mechanical resonant tunneling in standard scalar QFT, at least if the tunneling has to satisfy a series of well motivated conditions. 


The precise way to bridge the gap between quantum field theory and quantum mechanics is to 
picture quantum field theory as infinite-dimensional quantum mechanics following the pioneering work of Banks, 
Bender and Wu \cite{Banks:1973ps}. We use their notion of the 
``most probably escape path'' (MPEP), the class of paths which dominate the path integral in the classically 
forbidden region. In particular quantum field theory can be reduced in the semi classical limit to quantum 
mechanics along a preferred  path in configuration space (the MPEP in the classically forbidden regions and the 
classical path in the classically allowed regions). Tunneling effects are controlled by a generalised potential 
(\ref{eq:Uphi}),  which may well bear very little resemblance to the original field theory potential, $V$. From 
this point we are able to address the issue of resonant 
tunneling in field theory. Given the fact that we have essentially reduced the problem to a quantum mechanical 
one, we might naively expect resonant tunneling to be possible for a suitably chosen generalised potential. 
However, recall that for resonance to occur, we also require the existence of an oscillating solution to the 
classical field equations satisfying a Bohr-Sommerfeld condition (\ref{bsft2}). If it exists, this solution 
would provide the  spring board for the field to tunnel between the false and true vacuum, with nearly unit 
probability. No such solution, satisfying appropriate boundary conditions, exists.  {\tony The absence of a suitable oscillating solution may come as a surprise when one considers the fact that after bubble nucleation the field inside the bubble oscillates around the local minimum of the potential. However, the surfaces upon which $\del_t \phi$ vanishes are hyperbolae\footnote{The existence of these hyperbolae allows us to interpret the bubble's interior as an open universe~\cite{open}.}, as opposed to surfaces of constant $t$, as required by  our fourth condition. This condition is crucial because it enables us to match the two MPEPs to the bound state, generating the complete resonant tunneling path.}

{\tony Since we never actually specify the relevant spatial volume, $\mathcal{V}$, it is easy to see that our theorem can be applied over any finite region. In particular, imagine a situation where there are three vacua, $\phi_{false}, ~\phi_{middle}$ and $~\phi_{true}$, with $V(\phi_{false})>V(\phi_{middle})>V(\phi_{true})$, and assume that a bubble of $\phi_{middle}$ has nucleated inside a region of $\phi_{false}$. Of course, the probability of nucleation is exponentially suppressed, but once it {\it has} nucleated, we can ask whether or not resonant tunneling can happen locally within the bubble's interior. To see that it cannot, simply choose $\mathcal{V}$ to be contained entirely within the bubble, and define $\phi_{middle}$ to be our new false vacuum, and then reapply our theorem. Of course, standard Coleman tunneling~\cite{Coleman} could certainly occur inside the bubble via the nucleation of a second bubble of $\phi_{true}$, but this  process is exponentially suppressed and has nothing to do with {\it resonant} tunneling. Although our theorem rules out resonant tunneling, it does not apply to standard Coleman tunneling, since this does not require the existence of the intermediate bound state. }

At the end of the day, does this mean {\it resonant tunneling} can never occur in quantum field theory? {\tony Certainly we have said nothing about non-standard scalar field theories, or indeed gauge fields. However, at least for standard scalar QFT, for resonance} to occur, we believe one would need to go beyond the WKB approximation (but that is beyond 
the scope of this paper), or else find a satisfactory way of evading our no-go theorem. We might be able to achieve 
this by altering the boundary conditions in our analysis. In~\cite{ruth}, tunneling processes with nonvanishing 
momenta at the transition points were considered. Vanishing momenta is usually required for continuity, but one 
may consider relaxing it here to see whether it alters our result. Another alternative would be to change the 
spatial  boundary conditions at $\partial \mathcal V$. eg by allowing for a path which has  $\phi=\phi_0 \neq 0$  
everywhere in $\mathcal{V}$, where  also $V(\phi_0)=0$. Tunneling processes in which the entire space tunnels at 
once certainly exist in gravitating field theories~\cite{HM}, but this was not included in our analysis. 
{\tony One may also consider tunneling in situations where the spatial sections are compact, then the results of
simulations using the Hartree approximation \cite{Baacke:2005gq} suggest that resonant tunneling can occur.}
Of course, 
there may well be other efficient tunneling mechanisms  in field theory that bear no resemblance whatsoever to 
resonant tunneling in quantum mechanics, such as DBI tunneling~\cite{shiu}. Another example was recently discussed by Gleiser 
{\it et al}~\cite{bubbling}. By switching on a finite temperature, they generated an initial state containing 
a finite density of oscillons, which collide to form critical bubbles of true vacuum. This ``resonant nucleation'' 
of critical bubbles is qualitatively very different to the resonant tunneling phenomenon we have discussed in this 
paper, but may be relevant in the early universe. 

\vskip 1cm
{\it Note added: Whilst this paper was in the final stages of completion, another paper~\cite{shiu} appeared, which contains significant overlap with section~\ref{WKBQFT}.} 
\acknowledgments We thank Laurence Eaves for explaining the solid-state version of tunneling. We would also like to thank Henry Tye, Gary Shiu and Saswat Sarangi for interesting discussions. PMS is supported
by STFC.

\vskip 1cm
\appendix{\noindent\Large \bf Appendices}
\section{WKB connection formulae}
\label{appendixA}

\FIGURE{\includegraphics[width=8cm, height=8cm]{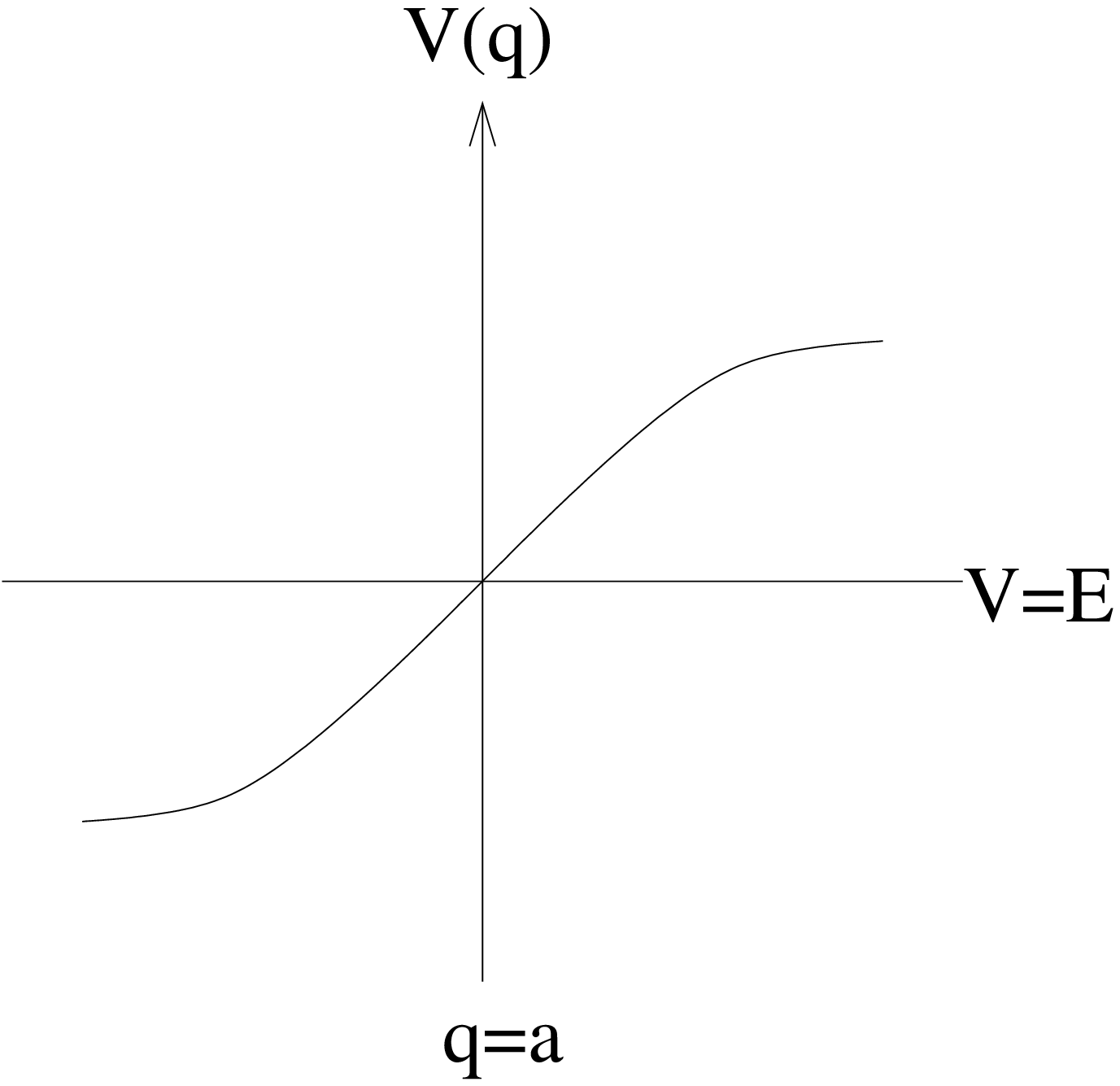}\includegraphics[width=8cm, height=8cm]{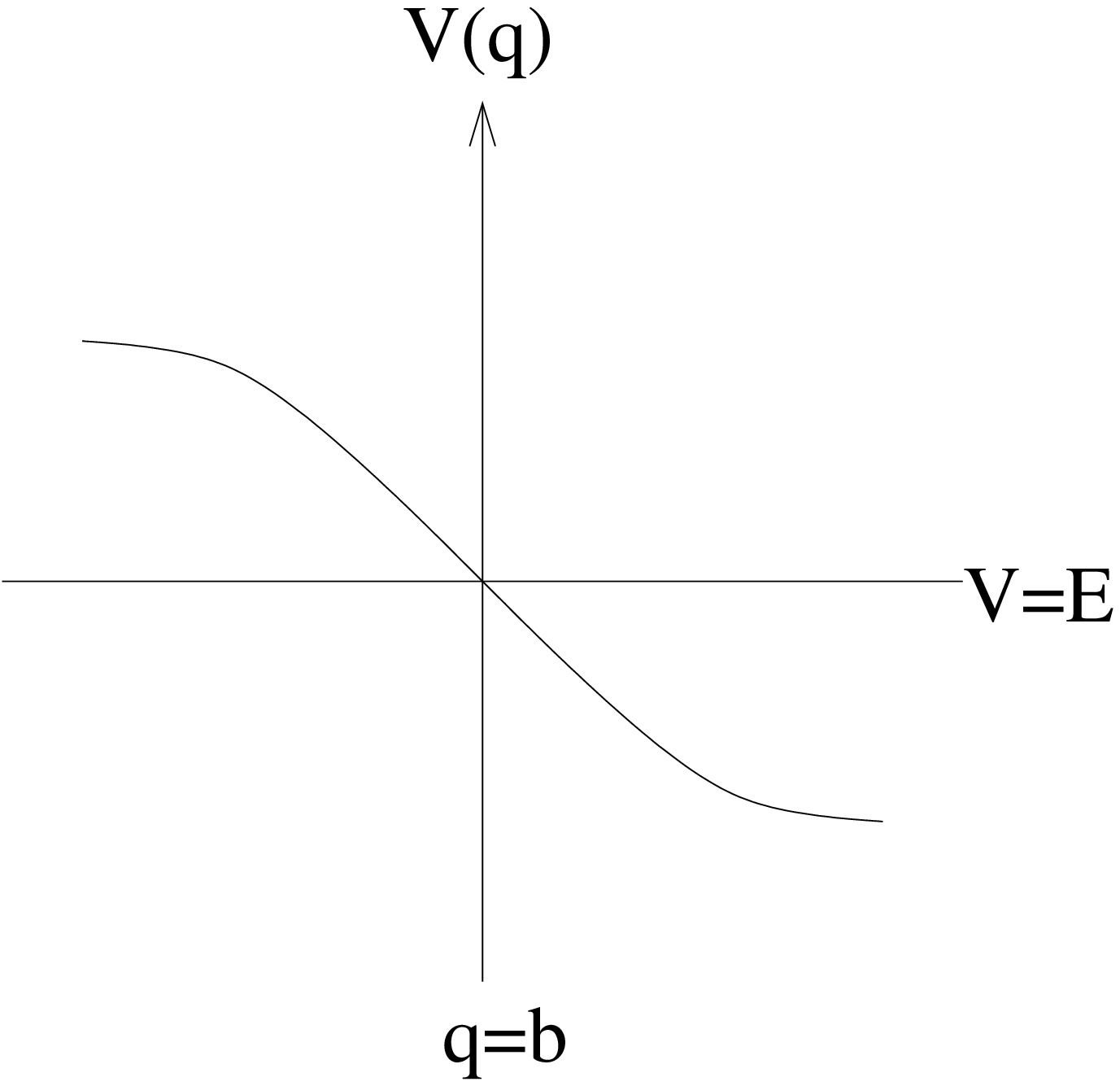}
\caption{Connecting the wavefunction when the potential increases (a) to the
         right of the turning point, or (b) to the left of the turning point.} \lab{fig:WKB}}

One has to be careful when using the WKB 
approximation to connect the wavefunction on either side of a classical
turning point.The reason being that in the classically forbidden region the solutions
are either exponentially growing or decaying, and in the presence of a growing solution the
WKB approximation is unable to track the decaying solution.
In practise what this means is that the connection
formulae depend on which direction you apply them. 

For example if we start in the classically
allowed region with the oscillating wavefunction then we must match this to the exponentially
growing solution, as the decaying solution will eventually be swamped by any amount of the growing
solution. The corrollary of this is that if we are given the growing solution and
try to attach the oscillating solution then we run into difficulties because the decaying
solution which the approximation misses effects the phase in the oscillating region.

One finds that when the potential increases to the right of the turning point
as in Fig. \ref{fig:WKB}(a) we connect according to \cite{llqm}
\ba
\label{eq:WKB1}
\frac{2}{\sqrt{k(q)}}\cos\left[\frac{1}{\hbar}\int^a_qk(q'){\rm d}q'-\pi/4\right]
&\leftarrow&\frac{1}{\kappa(q)}\exp\left[-\frac{1}{\hbar}\int_a^q\kappa(q'){\rm d}q'\right],\\
\label{eq:WKB2}
\frac{1}{\sqrt{k(q)}}\cos\left[\frac{1}{\hbar}\int^a_qk(q'){\rm d}q'+\alpha\right]
&\rightarrow&\frac{1}{\kappa(q)}\exp\left[\frac{1}{\hbar}\int_a^q\kappa(q'){\rm d}q'\right].
\ea
The notation means that we follow the arrow from the known solution to give us the
matched solution on the other side of the turning point. For example, (\ref{eq:WKB1}) tells
us that if we know the solution is exponentially decaying to the right in Fig. \ref{fig:WKB}(a)
then on the left we find the oscillating wavefunction of (\ref{eq:WKB1}). Similarly,
if we know the oscillatory wavefunction to the left of Fig. \ref{fig:WKB}(a) then for a genenric
phase $\alpha\neq-\pi/4$ we will get the growing mode given in (\ref{eq:WKB2}), which would swamp any decaying mode.

If we have a turning point where the potential increases to the left then onefinds the
following connection formulae,
\ba
\label{eq:WKB3}
\frac{1}{\kappa(q)}\exp\left[-\frac{1}{\hbar}\int_q^b\kappa(q'){\rm d}q'\right]
&\rightarrow&\frac{2}{\sqrt{k(q)}}\cos\left[\frac{1}{\hbar}\int^q_b k(q'){\rm d}q'-\pi/4\right],\\
\label{eq:WKB4}
\frac{1}{\kappa(q)}\exp\left[\frac{1}{\hbar}\int_q^b\kappa(q'){\rm d}q'\right]
&\leftarrow&\frac{1}{\sqrt{k(q)}}\cos\left[\frac{1}{\hbar}\int^q_bk(q'){\rm d}q'+\alpha\right].
\ea

\end{document}